\documentclass[showpacs,preprintnumbers,amsmath,amssymb]{revtex4}
\usepackage{amsmath}
\usepackage{graphicx}
\usepackage{subfig}
\usepackage{hyperref}
\usepackage{amssymb}
\usepackage[english]{babel}
\usepackage{epsfig}
\usepackage{wasysym}

\newcommand{\R}{\mathbb{R}}
\newcommand{\Hc}{\mathcal{H}}

\begin{document}
\title{Quantum cosmology of scalar-tensor theories and self-adjointness\\}
\vspace{0.3cm}
\author{Carla R. Almeida$^{(a)}$}\email{carlagbjj@hotmail.com}
\author{Antonio B. Batista$^{(a)}$}\email{abrasilb918@gmail.com}
\author{J\'{u}lio C. Fabris$^{(a,d)}$}\email{fabris@pq.cnpq.br}
\author{Paulo V. Moniz$^{(c,e)}$}\email{pmoniz@ubi.pt}
\vspace{0.5cm}

\bigskip

\affiliation{$^{(a)}$ Universidade Federal do Esp\'{\i}rito Santo, CEP 29075-910, Vit\'{o}ria/ES, Brazil}

\affiliation{$^{(c)}$ Departamento de F\'{\i}sica, Universidade da Beira Interior,
Covilh\~a, 6200, Portugal}

\affiliation{$^{(d)}$ National Research Nuclear University “MEPhI”, Kashirskoe sh. 31, Moscow 115409, Russia}

\affiliation{$^{(e)}$ Centro de Matem\'atica e Aplica\c{c}\~oes , CMA-UBI, Universidade da Beira Interior,
Covilh\~a, 6200, Portugal}

\begin{abstract}

In this paper, the problem of the self-adjointness for the case of a quantum minisuperspace Hamiltonian retrieved from a Brans-Dicke (BD) action is investigated. Our matter content is presented in terms of a perfect fluid, onto which the Schutz's formalism will be applied. We use the von Neumann theorem and the similarity with the Laplacian operator in one of the variables to determine the cases where the Hamiltonian is self-adjoint and if it admits self-adjoint extensions. For the latter, we study which extension is physically more suitable.
\end{abstract}
\pacs{ 04.50.Kd}

\maketitle

\section{Introduction}

General Relativity (GR) theory, the modern theory of gravity, has been totally successful in explaining the gravitational phenomena occurring at the solar system level, as well as in astrophysical systems in the strong gravitational regime, namely pulsar binary and compact stellar objects. At a cosmological level, it is the basic pillar in models allowing to predict the primordial abundance of light chemical elements and a relic of the hot radiative phase, the cosmic microwave background. These striking successes of the GR theory, however, do not hide its problems. More precisely, the necessity of a dark sector, responsible for 95\% of the matter-energy of the universe, and the existence of geometric singularities in the beginning of the history of the universe, as well as in the end stage of the life of some massive stars. These are indications that seem to point, at cosmological and astrophysical levels, to the limit of applicability of the GR theory.

The universality of quantum mechanics would imply that the GR theory must be quantised. If such quantisation of the gravitational theory is possible in a consistent way, it might be possible to solve the singularity problems in the strong gravitational regime. There is a lot of progress in the program of quantising gravity, either using canonical methods \cite{wdw1,wdw2}, or some different approaches, like loop quantisation \cite{loop1,loop2} or string theory
\cite{string1,string2}. But, it is yet too early to state that we possess a full quantum gravity theory. On the other hand, a simplified approach, like the quantisation of the Einstein-Hilbert action in the mini-superspace, in presence of matter fields, reveals that it is possible to obtain cosmological models free of singularities. This gives a strong support to the expectation that the singularity problems of the GR theory may be solved by evoking quantum phenomena \cite{kiefer}.

The construction of a quantum cosmological model faces many challenges, even when the mini-superspace restriction is used. The first one is the absence of an explicit time coordinate, due to the invariance by time reparametrisations in the classical theory \cite{tempo1,tempo2}. There are many different ways to cope with this problem. One of them is to allow the matter fields to play the r\^ole of time. In Refs. \cite{demaret,rubakov,lemos1,lemos2,nelson1} the implementation of the time coordinate at quantum level was done through the Schutz's method \cite{schutz}, whose canonical formulation allows to recover a genuine Schr\"odinger-type equation, because the conjugate momentum associated to the matter variables appears linearly in the Hamiltonian.

The second problem is the use of an appropriate interpretation formalism in quantum theory to obtain specific predictions. This is a very delicate problem, since the usual Copenhagen interpretation of quantum mechanics can not be translated in a direct way to quantum cosmology: the Copenhagen interpretation is based in a probabilistic formalism, using concepts as decoherence and a measurement mechanism through the spectral theorem, and it is not suitable for a system consisting of a unique realisation as the universe. However, many adaptations of the Copenhagen interpretation seem to be possible, as it happens with the Many World \cite{mw} or the Consistent Histories \cite{consistent}. Another less orthodox formulation is the Bohm-de Broglie (BdB) interpretation of quantum mechanics \cite{bdb1,bdb2}, which keeps the concept of trajectories of a given system, and does not use the probabilistic scheme\cite{nelson2}.

In ordinary quantum mechanics a central concept is the notion of self-adjoint operators: self-adjoint operators are essentially connected with the definition of observables and to the construction of the Hilbert space related to them. Hence, self-adjoint operators define in some sense the physical content of a given problem. To which extend such essential notion plays a r\^ole in quantum cosmology, where there is just a single system which contains the observer? We can expect that the answer to such question depends on the interpretation framework. But we can also expect that the specific predictions resulting from a given quantum cosmological model may depend on the character of the effective Hamiltonian operator and the boundary conditions related to this operator. In this case, it may depend on the self-adjointness property of the effective Hamiltonian operator and its possible self-adjoint extensions, which could fix the boundary conditions. 

In the context of GR theory, in the mini-superspace scheme, with the time coordinate induced by a perfect fluid described through Schutz's framework, the self-adjoint properties of the effective Hamiltonian has been investigated in Refs. \cite{lemos1,lemos2} for the case of isotropic and homogenous space-times. 
In such case, the quantum system is one-dimensional depending on the ordering factor only. These results have been extended to anisotropic spaces in Refs. \cite{pal1,pal2}. The case of the GR theory with a minimally coupled scalar field and ordinary matter has been treated in Ref. \cite{Ours}. In all these cases, which has a more involved phase space, it was necessary to address the problem of factor ordering and of the measure of the inner product, to obtain a consistent definition of self-adjoint Hamiltonian operator. Moreover, in Ref. \cite{Ours} it has been shown that the predictions to the evolution for the universe from a quantum cosmological model does not change with a different choice of ordering and the eigenfunctions of the Hamiltonian satisfy the boundary conditions even when we have extensions. In other words, the physical predictions depend very weakly on the self-adjoint character of the effective Hamiltonian operator. 

The objective of the present work is to extend these studies to scalar-tensor theories. We will consider the prototype of the scalar-tensor theories, the Brans-Dicke gravity theory, which differs from the case treated in \cite{Ours} by a conformal transformation. We will investigate if this kind of transformation interfers in any way with the self-adjoint aspect of the theory. In other words, the purpose will be the determination of the self-adjoint properties of the effective Hamiltonian operator in the mini-superspace, with the time variable conveyed by a perfect fluid described by the Schutz's procedure (for other possible approach to the time coordinate problem in the BD theory, see Ref. \cite{Valiki}). When the Hamiltonian is not self-adjoint but admits several extensions, as each extension describe a different physical settings \cite{R-S}, we suggest the extension that seems more natural mathematicaly, in terms of boundary conditions \cite{Bulla-Gesztesy}, with the intention of not to restrain the physical predictions too much. The impact of the results here obtained for specific predictions for the evolution of the universe are postponed to a future analysis.

This paper is organised as follows. In next section we quantise a model with a Lagrangian with a non-minimal coupled scalar field and matter fluid, establishing the corresponding Wheeler-DeWitt equation in the mini-superspace. In section III, we study the conditions for the resulting Hamiltonian operator to be symmetric in its domain. In section IV we determine the conditions for it to be self-adjoint or if it has any self-adjoint extensions, analysing two different particular cases which allows an analytical treatment: with radiative and stiff matter. We also study the extensions properties, in order to determine which one to choose. Our final remarks are presented in section V.

\section{Quantisation of the Brans-Dicke theory with non-minimal coupling}

For our purpose, let us consider the Lagrangian of a scalar field 
non-minimally coupled to gravity plus a generic matter component:
\begin{equation}
\mathcal{L}=\sqrt{-g}\left\{ \varphi R-\omega \frac{\varphi _{;\rho }\varphi
^{;\rho }}{\varphi }\right\} +\mathcal{L}_{M}\; .
\label{Lagrangian}
\end{equation}%
The gravitational Lagrangian defines the Brans-Dicke theory. 
Using a flat Friedmann-Lema\^{\i}tre-Robertson-Walker (FLRW) metric,
\begin{eqnarray}
ds^{2}=N\left( t\right) ^{2}dt^{2}-a\left( t\right)^{2} \left[dx^{2}+dy^{2}+dz^{2}\right] \quad,
\label{metric FLRW}
\end{eqnarray}
and discarding the surface terms, the Lagrangian can be written as,
\begin{equation}
\mathcal{L}_{G}=\frac{1}{N}\left\{ 6\left[ \varphi a\dot{a}^{2}+a^{2}\dot{a}%
\dot{\varphi}\right] -\omega a^{3}\frac{\dot{\varphi}^{2}}{\varphi }\right\} \quad,
\label{Lagrangian 2}
\end{equation}%
It can be written as function of the conjugated momenta, defined by
\begin{eqnarray}
\pi_q = \frac{\partial \mathcal{L}}{\partial \dot q} \quad,
\label{momentum}
\end{eqnarray} 
to obtain the Hamiltonian of the system, by using a Legendre transformation,
\begin{equation}
\mathcal{H}=N\left\{ \frac{1}{\left( 3+2\omega \right) }\left[ \frac{\omega 
}{12\varphi a}\pi _{a}^{2}+\frac{1}{2a^{2}}\pi _{a}\pi _{\varphi }-\frac{%
\varphi }{2a^{3}}\pi _{\varphi }^{2}\right] -\frac{1}{a^{3\alpha }}\pi
_{T}\right\} \quad.
\label{Total Hamillt.}
\end{equation}%
In this equation, $\pi_T$, $\pi _{a}$ and $\pi _{\varphi }$ are the conjugate
momenta associated with the matter component, the scale factor $a$ and the field $\varphi $, respectively. 

We have introduced the matter component, defined by the equation of state $p = \alpha\rho$, using Schutz's formalism \cite{lemos2}. In more detail, for a FLRW Universe, the fluid's four-velocity is written as \cite{schutz}
\begin{equation}
U_{\nu} = \frac{1}{\mu} \left( \epsilon_{;\mu} + \theta s_{;\mu} \right) \quad,
\label{four-velocity}
\end{equation}
with $\mu$ and $s$ being the specific enthalpy and entropy, respectively. The variables $\epsilon$ and $\theta$ do not have a clear physical meaning. The four-velocity must obey the normalization condition
\begin{equation}
U^{\nu}U_{\nu} = 1 \quad.
\label{normalization cond.}
\end{equation}
Using thermodynamical considerations that are described in details in Refs. \cite{demaret,rubakov,lemos1,lemos2,nelson1}, the Lagrangian of matter takes the form,
\begin{equation}
\mathcal{L}_{M} = - \frac{a^{3}}{N^{\frac{1}{\alpha}}} \frac{\alpha}{(\alpha + 1)^{\frac{1}{\alpha +1}}} \left( \dot{\epsilon} + \theta \dot{s} \right)^{\frac{1}{\alpha +1}} e^{- \frac{s}{a}} \quad, 
\label{Lagragiana de matéria}
\end{equation}
where we already discarded the surfaces terms. Using canonical methods \cite{rubakov}, from this Lagrangian we obtain the matter super-Hamiltonian
\begin{equation}
\mathcal{H}_{M} = - \pi_{\epsilon}^{\alpha +1} a^{-3 \alpha} e^{s} \quad, 
\label{super Hamiltoniana de matéria}
\end{equation} 
where $\pi_{\epsilon} = -N \rho_{0} U^{0} a^{3}$ and with $\rho_{0}$ being the rest mass density of the fluid. Using the canonical transformation \cite{rubakov,lemos1}
\begin{equation}
T= - \pi_{s} e^{-s} \pi_{\epsilon}^{-(\alpha +1)} \quad; \quad \pi_{T} = \pi_{\epsilon}^{\alpha +1} e^{s} \quad; \quad \overline{\epsilon} = \epsilon - (\alpha +1) \frac{\pi_{s}}{\pi_{\epsilon}} \quad; \quad \overline{\pi_{\epsilon}} = \pi_{\epsilon} \quad,
\label{transfor. canonicas}
\end{equation}
the super-Hamiltonian of the matter component becomes: 
\begin{equation}
\mathcal{H}_{M}=-\frac{N}{a^{3\alpha }}\pi _{T} \quad,
\label{Hamilt. matter}
\end{equation}%
as introduced in [\ref{Total Hamillt.}].

The quantisation of the equation ($\ref{Total Hamillt.}$), by means of a
Wheeler-DeWitt equation perspective, results in a Schr\"odinger-like equation, from where
we obtain a Hamiltonian operator. The canonical quantisation 
$\pi _{k}\rightarrow -i\partial _{k}$, including the ordering factors
$\bar{p}$ and $\bar{q}$, gives us thus,
\begin{equation}
\frac{\omega }{12\varphi }\left( \frac{1}{a}\partial _{a}^{2}\Psi +\frac{%
\bar{p}}{a^{2}}\partial _{a}\Psi \right) +\frac{1}{2a^{2}}\partial
_{a}\partial _{\varphi }\Psi -\frac{1}{2a^{3}}\left( \varphi \partial
_{\varphi }^{2}\Psi +\bar{q}\partial _{\varphi }\Psi \right) =\frac{i\left(
3+2\omega \right) }{a^{3\alpha }}\partial _{T}\Psi \quad.
\label{Schroed. 1}
\end{equation}%
In order to avoid the derivative $\partial _{a}\partial _{\varphi }$,
let us make a change of coordinates,
\begin{equation}
a\rightarrow \varphi ^{-\frac{1}{2}}b;\qquad \varphi \rightarrow \varphi \quad.
\label{coordinates transf.}
\end{equation}%
With these new coordinates and redefining $T\rightarrow T/24$, equation %
($\ref{Schroed. 1}$) becomes%
\begin{equation}
-\left\{ \frac{\varphi ^{\frac{1}{2}}}{b}\left[ \partial _{b}^{2}+\frac{p}{b}%
\partial _{b}\right] - \frac{12}{(3+2 \omega)}\frac{\varphi ^{\frac{3}{2}}}{b^{3}}\left[ \varphi
\partial _{\varphi }^{2}+q\partial _{\varphi }\right] \right\} \Psi =\frac{%
\varphi ^{\frac{3\alpha }{2}}}{b^{3\alpha }}i\partial _{T}\Psi \quad,
\label{Schroed. 2}
\end{equation}%
where $p$ and $q$ are new ordering factors. They are related to the previous ones in the following way:%
\begin{equation}
p=\frac{2\omega \bar{p}+9-6\bar{q}}{3+2\omega };\qquad q=\bar{q} \quad.
\label{p,q relations}
\end{equation}

We can verify that the Hamiltonian operator
\begin{equation}
\hat{H}=-\frac{b^{3\alpha }}{\varphi ^{\frac{3\alpha }{2}}}\left\{ \frac{%
\varphi ^{\frac{1}{2}}}{b}\left[ \partial _{b}^{2}+\frac{p}{b}\partial _{b}%
\right] - \frac{12}{(3+2\omega)} \frac{\varphi ^{\frac{3}{2}}}{b^{3}}\left[ \varphi \partial_{\varphi }^{2}+q\partial _{\varphi }\right] \right\} \quad,
\label{Hamiltonian operator}
\end{equation}%
coming from equation ($\ref{Schroed. 2}$), is symmetric on the domain
consisting of smooth compactly supported functions, i.e.,
\begin{equation}
\left\langle \phi ,H\psi \right\rangle = \left\langle H\phi ,\psi \right\rangle,
\quad  \phi,\psi\in C_0^\infty(\R_+^2) \quad,
\label{domain H}
\end{equation}
with the scalar product
\begin{equation}
\left\langle \phi ,\psi \right\rangle =\int_{0}^{\infty }\int_{0}^{\infty
}\left( \overline{\phi }\psi \right) b^{p-3\alpha +1}\varphi ^{q+\frac{%
\left( 3\alpha -5\right) }{2}}dbd\varphi \quad.  
\label{measure}
\end{equation}
Notice that the measure defining the  scalar product depends on the 
ordering factors. 

\section{Self-adjointness and self-adjoint extensions}

Usually, the notions of symmetric and self-adjoint operators are taken as synonymous, yet they have to be distinguished as only self-adjoint
operators give rise to a one-parameter unitary group (Stone theorem) \cite{R-S - vol1}. The self-adjointeness of an operator requires the condition that its domain coincides with the domain of its adjoint operator. In our case, since the Hamiltonian operator (\ref{Hamiltonian operator}) has real-valued functions as coeficients, then its adjoint has the same form as in (\ref{Hamiltonian operator}), the difference between the two of them is only with respect to the domain: the domain of the adjoint $\hat{H}^{*}$ may be bigger than the domain of $\hat{H}$. That is, $D(\hat{H}) \subset D(\hat{H^{*}})$. If $\hat{H}$ is symmetric \textit{and} $D(\hat{H}) = D(\hat{H^{*}})$, the operator is then self-adjoint. See \cite{R-S} for further details and examples.

 Nevertherless, if an operator is only symmetric (densely-defined) there are cases when its domain can be extended  in order to become self-adjoint, that is, the operator with this new domain has an adjoint operator with the same domain. This operator with the extended domain is called a self-adjoint extension. Not all densely-defined symmetric operators can be extended to a self-adjoint operator, but when it can be extended, it may have one or  infinitely many different self-adjoint extensions.  

The question of self-adjoint extensions of a symmetric operator has been thourougly studied by von Neumann shortly after Quantum Mechanics gradually emerged. An useful criterion for the existence of such extensions is given by a theorem by von Neumann \cite{R-S}:

\vspace{0.5cm}
\noindent
{\it Let $A$ be a symmetric operator with dense domain $D\left( A\right)$ in a Hilbert space $\Hc$, and suppose there exists an antilinear map $C:\Hc\rightarrow \Hc$ such that $C(D(A))\subset D(A)$, $C^{2}=Id$ ($C$ is called conjugation map) and $CA=AC$ on $D(A)$. Then $A$ admits self-adjoint extensions.}
\vspace{0.5cm}

In our case, the map $C$ would simply be the complex conjugation $\phi \mapsto \bar\phi$ in the Hilbert space  of square-integrable functions for the (positive) measure defined by (\ref{measure}):
\begin{equation}\
d\mu(b,\varphi)= b^{p-3\alpha +1}\varphi ^{q+\frac{\left( 3\alpha -5\right) }{2}}dbd\varphi \quad.
\label{mea}
\end{equation}
We shall consider the operator $\hat H$ given in ($\ref{Hamiltonian operator}$) with domain $D(\hat H) = C_0^\infty(\R_+^2)$, the space of smooth compactly supported functions $\phi(b,\varphi)$, where $0\leq b,\varphi<\infty$. Then, we have
\begin{equation}
C^{2}=Id;\qquad \quad C\hat{H}\phi=\hat{H}C\phi,\quad \phi \in D(\hat H) \quad,
\end{equation}%
because the Hamiltonian is a differential operator with real coefficients. Therefore the von Neumann theorem assures us that the Hamiltonian ($\ref{Hamiltonian operator}$) has self-adjoint extensions. 

Moreover, we can study self-adjoint extensions via deficiency indices. Given a symmetric operator $A$ with dense domain $D(A)$ let us denote by $K_\pm(A)$ the space of solutions $\phi$ of
\begin{equation}
A^*(\phi) = \pm i \phi,\quad \phi\in D(A^*) \quad.
\label{indices equation}
\end{equation}
The subspaces $K_\pm(A)$ are called the deficiency spaces and their dimensions $n_\pm =\rm{dim }K_\pm(A)$
are the deficiency indices. A basic theorem of von Neumann states that if $n_{+}= n_{-}= 0$ the operator has a unique self-adjoint extension, that is, it is essentially self-adjoint, if $n_{+}= n_{-} = n$ self-adjoint extensions are in one-to-one correspondance with unitary operators between $K_{+}$ and $K_{-}$, and if $n_{+} \neq n_{-}$ the operator does not admit any self-adjoint extensions. 

Essentially self-adjoint operators have a unique extension. But from now on, we will ignore this distinction with respect to self-adjoint operators in a strict sense and call both as self-adjoint operators. 
Moreover, for a positive symmetric operator, there is a unique self-adjoint extension, called the Friedrich's extension, that can be obtained from the quadratic form associated with the operator $A$ (Theorem X.23 from \cite{R-S}).

\vspace{0.5cm}
\noindent
{\it Let $A$ be a positive symmetric operator and let $q(\phi, \psi) =  \langle \phi, A\psi \rangle$ for $\phi,\psi \in D(A)$. Then $q$ is a closable quadratic form and its closure $\hat{q}$ is the quadratic form of a unique self-adjoint operator $\hat{A}$. $\hat{A}$ is a positive extension of $A$, and the lower bound of its spectrum is the lower bound of $q$. Further, $\hat{A}$ is the only self-adjoint extension of $A$ whose domain is contained in the form domain of $\hat{q}$.}
\vspace{0.5cm}

A quadratic form is a map $q: Q(q) \times Q(q) \rightarrow \mathbb{C}$ which is anti-linear on the first argument and linear on the second one. The set $Q(q)$ is called \textit{Form Domain} and it is a dense subset of the Hilbert space. A positive quadratic form, that is $q(\psi, \psi) \geq 0$ for $\psi \in Q(q)$, is said to be closed if $Q(q)$ is complete under the norm $\|\psi \|^{2}_{+1} \equiv \|\psi \|^{2} + q(\psi,\psi)$, which is called \textit{Form Norm}. 
For a symmetric and positive operator $A$, we can define the positive symmetric quadratic form
\begin{equation}
q(\phi, \psi) = \langle \phi, A\psi \rangle \quad,
\label{quadratic form of A}
\end{equation} 
with $Q(q) = D(A)$ being the domain of $A$. The closure $\hat{q}$ of $q$ is build using the Form Norm and, by the Theorem above, it is associated with a unique positive self-adjoint extension $\hat{A}$ such that $D(\hat{A}) \subset Q(\hat{q})$.

Mathematically, all the possible extensions are viable, and, physically, each of them generates different dynamics.  Nevertherless, it seems that the Friedrich's extension is the most natural one. In fact it always exists if the operator is bounded from below (in particular, if it is positive). Furthermore, it has the same lower bound as the operator and it is the only one whose domain is contained in the form domain of the quadratic form defined from the operator. In other words, it preserves the ground state energy, and the boundary conditions are more evident compared to other extensions, due to the fact that it can be obtained from the quadratic form of the operator. Nevertheless, it is worth to mention that, in our case, the Hamiltonian operator is not bounded from below if  $\varpi >0$, but it is positive and bounded if $\varpi < 0$.
As an example, for the operator $A = -d^2/dx^2$ on $C_{0}^{\infty}(0,1)$, the Friedrich's extension $\hat{A}$ is given by the boundary conditions $\psi(0) = 0 = \psi(1)$, where $\psi \in D(\hat{A})$. \cite{R-S}

These nice properties and uniqueness of the Friedrich's extension is an advantage and it is adopted in many diferent contexts, such as anisotropic models \cite{pal3}, or to obtain a unitary time evolution imposing Friedrich's boundary conditions on a  Reissner-Norsdstr\"om singularity \cite{black hole}, or in more general theories \cite{Gazeau}.

\section{Examples: radiative and stiff matter}

So far, we know that the Hamiltonian operator \ref{Hamiltonian operator} has self-adjoint extensions and, when it is positive (i.e. when $\varpi <0$), it admits a positive self-adjoint extension, namely the Friedrich's extension. We will show that it depends on the ordering factors $p$ and $q$, as well as on the constant $\varpi$. For this, let us study the self-adjoint extensions of $\hat H$ determining its deficiency indices $n_{\pm }$, which corresponds to the number of independent (distribution) solutions in the domain $D(\hat{H^{*}}) = \{\psi \in L^{2} / \hat{H} \psi \in L^{2}\}$ of the equation 
\begin{equation}
\hat{H}\Psi =\pm i\Psi \quad,
\label{eigenvalue equation}
\end{equation}%
where $L^{2}$ is the set of functions that are square-integrable for the measure (\ref{mea}). Using the operator ($\ref{Hamiltonian operator}$), Eq. (\ref{eigenvalue equation}) takes the explicity form:
\begin{equation}
-\frac{b^{3\alpha }}{\varphi ^{\frac{3\alpha }{2}}}\left\{ \frac{\varphi ^{%
\frac{1}{2}}}{b}\left[ \partial _{b}^{2}+\frac{p}{b}\partial _{b}\right] -\varpi%
\frac{\varphi ^{\frac{3}{2}}}{b^{3}}\left[ \varphi \partial _{\varphi
}^{2}+q\partial _{\varphi }\right] \right\} \Psi = -\eta \Psi \quad,
\end{equation}
or, equivalently
\begin{equation}
\frac{\varphi ^{\frac{1}{2}}}{b}\left[ \partial _{b}^{2}+\frac{p}{b}\partial
_{b}\right] \Psi -\varpi \frac{\varphi ^{\frac{3}{2}}}{b^{3}}\left[ \varphi
\partial _{\varphi }^{2}+q\partial _{\varphi }\right] \Psi = -\eta \frac{%
\varphi ^{\frac{3\alpha }{2}}}{b^{3\alpha }}\Psi \quad,
\end{equation}
with $\eta =\pm i$ and $\varpi = 12(3 + 2\omega)^{-1}$. However, if we impose 
$\Psi \left( b,\varphi \right) =X\left( b\right) Y\left( \varphi \right) $ we obtain the following partial differential equation:
\begin{equation}
b^{2}\left[ \frac{\ddot{X}}{X}+\frac{p}{b}\frac{\dot{X}}{X}\right]
-\varpi \varphi \left[ \varphi \frac{Y^{\prime \prime }}{Y}+q\frac{Y^{\prime }}{Y}%
\right] =-\eta \frac{\varphi ^{\frac{3\alpha -1}{2}}}{b^{3\left( \alpha
-1\right) }} \quad,  
\label{PDE}
\end{equation}%
where an over dot denotes the derivation with respect to $b$ and the prime is the derivation 
with respect to $\varphi $. This equation is separable only for the particular 
cases of $\alpha =\frac{1}{3}$ or $\alpha =1$, that is, only for radiative and stiff matter. Therefore, for these cases, the set of square-integrable functions with the measure (\ref{mea}) is a direct product given by
\begin{equation}
L^{2}(\mathbb{R}^{2}_{+}, d\mu (b, \varphi)) = L^{2}(\mathbb{R}_{+}, b^{p-3\alpha + 1}db) \otimes L^{2}(\mathbb{R}_{+}, \varphi ^{q + \frac{(3\alpha - 5)}{2}}d\varphi) \quad.
\end{equation}  
This means that we can separate the domain of the Hamiltonian into a direct product $D(\hat{H}) = D(\hat{H}_{b}) \otimes D(\hat{H}_{\varphi})$ where $\hat{H}_{b}$ and $\hat{H}_{\varphi}$ are the operators that compose the Hamiltonian, with respect to $b$ and $\varphi$, respectively. Using Theorem VIII.33 presented in \cite{R-S - vol1}, if $\hat{H}_{b}$ and $\hat{H}_{\varphi}$ are self-adjoint, then $\hat{H}$ is self-adjoint. Thus we will use the separation variable method for these two cases.

\subsection{First case: $\protect\alpha =\frac{1}{3}$}
\label{Subsec 1/3}

For this case, operator (\ref{Hamiltonian operator}) becomes
\begin{equation}
\hat{H} = - \partial_{b}^{2} - \frac{p}{b} \partial_{b} + \frac{\varpi}{b^{2}} \left[ \varphi^{2} \partial_{\varphi}^{2} + q\varphi \partial_{\varphi} \right] \quad.
\label{Hamilt. operator 1/3}
\end{equation}
When $\alpha = 1/3$, the fluid is conformal invariant at classical level. We may expect that this property remains at quantum level, and the results of Ref. \cite{Ours} must be recovered. We will verify this explicitly.
Consider the unitary operator $U: L^{2}( \mathbb{R}_{+}^{2}, b^{p} \varphi^{q-2} dbd\varphi) \rightarrow L^{2}( \mathbb{R}_{+}^{2},\varphi^{q-2} dbd\varphi)$ such as $U: \phi (b, \varphi) \mapsto b^{p/2} \phi (b, \varphi)$. $U$ takes the domain $D(\hat H) = C_c^\infty(\R_+^2)$ into itself, preserving the properties of the operator $\hat{H}$. Thus: 
\begin{eqnarray}
U\hat{H}U^{-1}= b^{\frac{p}{2}} \left\{ - \partial_{b}^{2} - \frac{p}{b} \partial_{b} + \frac{\varpi}{b^{2}} \left[ \varphi^{2} \partial_{\varphi}^{2} + q\varphi \partial_{\varphi} \right] \right\}  b^{-\frac{p}{2}} 
\nonumber\\
= - \partial_{b}^{2} + \frac{1}{b^2} \left\{ \frac{p}{2} \left(\frac{p}{2} -1 \right) + \varpi \left[ \varphi^{2} \partial_{\varphi}^{2} + q\varphi \partial_{\varphi} \right] \right\} \quad.
\label{example S-R case 1/3}
\end{eqnarray}
Therefore, from the eigenvalue equation (\ref{eigenvalue equation}), we obtain:
\begin{equation}
 \hat{H}^{'} \Psi^{'} = \eta \Psi^{'} \quad, 
 \label{eigenvalue eq. mod.}
\end{equation}
where $ \hat{H}^{'} = U \hat{H} U^{-1}$ and $\Psi^{'} = U\Psi$. 
Using separation of variables, equation (\ref{eigenvalue eq. mod.}) becomes a system of differential equations:
\begin{eqnarray}
\hat{H}_{b} X(b) = \left\{ - \partial_{b}^{2} + \frac{1}{b^2} \left[ \frac{p}{2} \left(\frac{p}{2} -1 \right) + \epsilon k^{2} \right] \right\} X(b) = \eta X(b) \quad, 
\label{OED b case 1/3}
\\
\hat{H}_{\varphi} Y(\varphi) = \varpi \left[ \varphi^{2} \partial_{\varphi}^{2} + q\varphi \partial_{\varphi} \right] Y(\varphi) = \epsilon k^{2} Y(\varphi) \quad,
\label{OED varphi case 1/3}
\end{eqnarray}
with $k \in \mathbb{R}$ and $\epsilon = \pm1$. Notice that equation \eqref{OED varphi case 1/3} remains the same for the calculation of the stationary states of energy. Hence, even in the analysis of the self-adjoint character of the Hamiltonian, following von Neumann method, we must require the function $Y(\varphi)$ to be square-integrable in order to have a physically well posed problem. We will see that a necessary condition for having square-integrability is that $q=1$. This means that the domain of operator $\hat{H}_{\varphi}$ is $L^{2}(\mathbb{R}_{+}, \varphi^{q-2} d\varphi)$ only for $q=1$, therefore, we will not consider the case $q\neq 1$.
	
To proceed with the von Neumann method to find the deficience indices, one must solve these equations and verify if the solutions are in the domain of the adjoint operator. However, we will consider the fact that $\hat{H}_{b}$ is a Laplacian-like operator,
\begin{equation}
\hat{H}_{b} = - \partial_{b}^{2} + \frac{1}{b^2} \left[ \frac{p}{2} \left(\frac{p}{2} -1 \right) + \epsilon k^{2} \right] \quad,
\label{Hb case 1/3}
\end{equation}
which is well known in the literature (see \cite{R-S}), to ensure a condition for its self-adjointness, using the following theorem:

\vspace{0.5cm}
\noindent
{\it Let $V(r)$ be a continuous symmetric potential on $\mathbb{R}_{+}|\{0\}$, then $-\Delta + V(r)$, with $\Delta$ being the Laplacian operator}
\begin{equation}
\Delta = \frac{\partial^{2}}{\partial_{r}^{2}} - \frac{\lambda}{r^2}\quad,
\label{laplacian operator}
\end{equation}
{\it will be essentially self-adjoint in $C_{0}^{\infty}$ only if $V(r)$ satisfies}
\begin{equation}
V(r) + \frac{\lambda}{r^2} \geq \frac{3}{4} \frac{1}{r^2} \quad.
\label{condition self-adjointness} 
\end{equation}
\textit{where $\lambda$ is a real constant.}
\vspace{0.5cm}

Therefore, in our case, operator (\ref{Hb case 1/3})  is essentially self-adjoint if
\begin{equation}
\frac{p}{2} \left(\frac{p}{2} -1 \right) + \epsilon k^{2} \geq \frac{3}{4} \quad \Rightarrow \quad \epsilon k^{2} \geq \frac{3}{4} - \frac{p}{2} \left(\frac{p}{2} -1\right) \quad \Rightarrow 
\begin{cases}
2 |k| \geq \sqrt{-p^{2} + 2p + 3} \quad; \, \mathrm{for} \quad \epsilon =1
\\
2 |k| \leq \sqrt{p^{2} - 2p  - 3} \quad;\,  \mathrm{for} \quad \epsilon =- 1
\end{cases}
\quad.
\end{equation}
For both intervals to be satisfied the polynomial inside the square root must be greater than or equal zero, hence we must have $-1 \leq p \leq 3$ for $\epsilon=1$ and $p\leq-1$ and $p\geq3$ for $\epsilon=-1$. By means of this, we established a condition over the ordering factor $p$ for the Hamiltonian operator to be self-adjoint. 

With this, we guarantee that there is at least one solution $X(b)$ of equation (\ref{OED b case 1/3}) for $p < -1$ and $p > 3$ if $\epsilon=1$ or for $-1<p<3$ if $\epsilon=-1$ such that $X \in D(\hat{H}_{b}) = C_{0}^{\infty} (\mathbb{R}_{+})$. Actually, the solutions are a combination of polinomials and Bessel functions, but we do not need to explicit them, since our goal is not to find how many solutions there are, but to determine the conditions for their existence. Let us refer to them as $X^{(i)}(b,k)$, as they depend on the separation constant $k$. Now, we must find out for which values of $q$ there are square-integrable solutions $Y(\varphi)$ of (\ref{OED varphi case 1/3}).

Now, let us verify that $q$ must be equal to $1$ indeed. Equation (\ref{OED varphi case 1/3}) it is a second-order Euler equation and the independent solutions are
\begin{equation}
Y_{(+)}\left( \varphi \right) =\varphi ^{\sigma + \rho}   \quad, \quad Y_{(-)}\left( \varphi \right) =\varphi ^{\sigma - \rho}
\label{Y case 1/3}
\end{equation}%
where we have, for a general $q$,
\begin{equation}
\sigma =\frac{1-q}{2};\qquad \rho =\sqrt{\sigma ^{2}+\epsilon \frac{k^{2}}{\varpi}} \quad.
\label{sigma, ro relations}
\end{equation}
With this, the general solutions of the eigenvalue equation in this case have the form
\begin{equation}
\Psi_{\pm}^{(i)}(b,\varphi) = \int_{\mathcal{K}} A(k) X^{(i)}(b,k) \varphi ^{\sigma \pm \rho} dk \quad.
\label{general solution case 1/3}
\end{equation}
The $A(k)$ is a normalization function and it has support compact in the interval $\mathcal{K} \subset \mathbb{R}$ where $k$ is validated. This equation is the reason why we have chosen the constant of separation in equations \eqref{OED b case 1/3} and \eqref{OED varphi case 1/3} to be strictly postive or strictly negative, because otherwise the wavepacket would diverge even in the physical eigenvalue case.

Let us verify which of the solutions are in the domain of the operator $D(\hat{H}^{*})$. Since they are eigenfunctions of $\hat{H}$, we only need to determine if they are square-integrable in the measure adopted. 
\begin{equation}
\left\Vert \Psi^{(i)}_{\pm} \right\Vert^{2} =\int_{0}^{\infty} \int_{0}^{\infty} \int_{\mathcal{K}} \int_{\mathcal{K}} \overline{A(k')}A(k) \overline{X^{(i)}(b,k')}X^{(i)}(b,k) b \varphi ^{\pm \left( \rho +\overline{\rho^{\prime }}\right) -1} dkdk'dbd\varphi
\label{norm wave funcion}
\end{equation}
where $\rho' = \rho(k')$. Changing the $\varphi $ coordinate as $\varphi \rightarrow e^{u}$, the integral over $\varphi$ becomes
\begin{equation}
\int_{0}^{\infty }\varphi ^{\pm \left( \rho +\overline{\rho }\right)
-1}d\varphi =\int_{-\infty }^{\infty }e^{\pm \left( \rho +\overline{\rho
^{\prime }}\right) u}du\quad .
\label{phi --> exp(u) int.}
\end{equation}
Notice that this integration diverges for both signs if we have $(\rho + \rho^{\prime})$ has a real component, that is, it must be imaginary only. Hence, tanking into consideration equations (\ref{sigma, ro relations}), for $(\rho + \rho^{\prime})$ to be imaginary we must have $\sigma = 0$ (that is $q = 1$), and $\varpi<0$ for $\epsilon=1$ or $\varpi>0$ for $\epsilon=-1$, with this we obtain convergent solutions. In these cases, we have $\rho =i\left\vert
k\right\vert /\sqrt{ \vert \varpi \vert}$ the integration over $u$ results in a Dirac's delta
function:
\begin{equation}
\int_{-\infty }^{\infty }\exp \left[ \pm i\frac{\left[ \left\vert
k\right\vert -\left\vert k^{\prime }\right\vert \right] }{\sqrt{\vert \varpi \vert}}u\right] du= 2\pi \sqrt{\vert \varpi \vert}\delta \left( \pm \left\vert k\right\vert \mp \left\vert k^{\prime }\right\vert \right)\quad .
\label{integ. over u}
\end{equation}%
This is the condition over $q$ that we were looking for and then equation \eqref{norm wave funcion} becomes
\begin{equation}
\left\Vert \Psi^{(i)}_{\pm} \right\Vert^{2}_{D(\hat{H})} = \int_{\mathcal{K}} \left\Vert A(k)X^{(i)}(b,k) \right\Vert^{2}_{D(\hat{H}_{b})} dk \quad.
\end{equation}

We have found the conditions for the existence of square-integrable solutions of equations (\ref{eigenvalue eq. mod.}) in terms of $p,q$ and $\varpi$. If those conditions are not satisfied, then there are no square-integrable functions that satisfy (\ref{eigenvalue eq. mod.}) and therefore the deficiency indices $n_{+}$ and $n_{-}$ are both null. The overall conclusion is:
\begin{enumerate}
	\item[1.] If $q=1$ and $\varpi > 0$, then there are two possibilities:
	\begin{enumerate}
			\item[i.]  For $ p\leq -1$ and $p \geq 3$ the Hamiltonian ($\ref{Hamiltonian operator}$) is already self-adjoint.
			
			\item[ii.] It is not self-adjoint otherwise, that is $-1 < p < 3$, thought it has self-adjoint extensions.
	\end{enumerate}
		
	\item[2.] If $q=1$ and $\varpi < 0$, we have similar possibilities:
	\begin{enumerate}
			\item[i.]  For $-1 \leq p \leq 3$ the Hamiltonian ($\ref{Hamiltonian operator}$) is already self-adjoint.
			
			\item[ii.] For $p < -1$ or $p > 3$ it is not self-adjoint, but it has self-adjoint extensions.
	\end{enumerate}
\end{enumerate}

We did not calculate the deficiency indices explicitly for every case, and we did not intend to do it, even if in the Appendix \ref{Anexo A} we sketch how the analysis can be made through the von Neumann method. The goal was to determine the cases which the operator has self-adjoint extensions, and to do it we only need to verify if it has equal indices. Nevertherless, our calculation suggests that there are an infinity number of self-adjoint extensions, for the cases where the Hamiltonian operator is not already self-adjoint. The reason behind it is related with the arbitrariness of the superposition factor $A(k)$ introduced in equation (\ref{general solution case 1/3}).

As we said before, the Hamiltonian operator is positive and bounded from below for the case $\varpi <0$, then we can adopt the Friedrich's extension for the case 2.ii. Nevertherless, there is not a privileged extension for the case 1.ii, because we have an unbounded (from below) operator.

\subsection{Second case: $\protect\alpha =1$}

The case $\alpha = 1$ is very similar to the previous one. The Hamiltonian operator (\ref{Hamiltonian operator}) becomes
\begin{equation}
\hat{H} = \varpi \varphi \left\{\partial_{\varphi}^{2} + \frac{q}{\varphi} \partial_{\varphi} - \frac{1}{\varphi^{2}} \left[ \frac{b^{2}}{\varpi} \left( \partial_{b}^{2} + \frac{p}{b} \partial_{b} \right)\right] \right\} \quad,
\label{Hamilt. operator 1}
\end{equation}
and it is symmetric with the measure $b^{p-2} \varphi^{q-1} dbd\varphi$. Let us consider the associated operator $\hat{H}_{1}$ such as $\hat{H} = \varphi \hat{H}_{1}$. Notice that 
\begin{equation}
\int \hat{H} \varphi^{q-1} d\varphi = \int \varphi \hat{H}_{1} \varphi^{q-1} d\varphi = \int \hat{H}_{1} \varphi^{q} d\varphi \quad,
\label{int H = int H1}
\end{equation}
hence, if $\hat{H}_{1}$ is self-adjoint on the domain $D(\hat{H}) = C^{\infty}_{0}(\mathbb{R}_{+}, b^{p-2}  db) \otimes C^{\infty}_{0} (\mathbb{R}_{+}, \varphi^{q-1} d\varphi)$, then $\hat{H}$ is self-adjoint on the domain $D(\hat{H}_{1}) = C^{\infty}_{0}(\mathbb{R}_{+}, b^{p-2}  db) \otimes C^{\infty}_{0} (\mathbb{R}_{+}, \varphi^{q} d\varphi)$. Therefore, we can verify the self-adjointness of operator $\hat{H}_{1}$ instead, which is similar up to a multiplicative constant to the operator (\ref{Hamilt. operator 1/3}) from the case $\alpha = 1/3$, but with the variables inverted. Then, we can proceed as the case before. Consider the unitary operator $V: L^{2}( \mathbb{R}_{+}^{2}, b^{p-2} \varphi^{q} dbd\varphi) \rightarrow L^{2}( \mathbb{R}_{+}^{2}, b^{p-2} dbd\varphi)$ such as $U: \phi (b, \varphi) \mapsto \varphi^{p/2} \phi (b, \varphi)$. We have
\begin{equation}
V \hat{H}_{1} V^{-1} = \varpi \left\{ \partial_{\varphi}^{2} - \frac{1}{\varphi^{2}} \left[\frac{q}{2} \left( \frac{q}{2} -1 \right) +  \frac{b^{2}}{\varpi} \left( \partial_{b}^{2} + \frac{p}{b} \partial_{b} \right) \right] \right\} \quad.
\label{example S-R case 1}
\end{equation}
Considering separation of variables, we obtain the following system of differential equations:
\begin{eqnarray}
\hat{H}_{1b} X(b) = \frac{b^{2}}{\varpi} \left( \partial_{b}^{2} + \frac{p}{b} \partial_{b} \right) X(b) = \epsilon k^{2} X(b) \quad;
\label{ODE X case1}
\\
\hat{H}_{1\varphi} Y(\varphi) = \varpi \left\{ \partial_{\varphi}^{2} - \frac{1}{\varphi^{2}} \left[\frac{q}{2} \left( \frac{q}{2} -1 \right) + \epsilon k^{2} \right] \right\} Y(\varphi) = \eta Y(\varphi) \quad.
\end{eqnarray}
For the same reasons as before, $p$ must be fixed equal to one, and the operator
\begin{equation}
\hat{H}_{1\varphi} = \varpi \left\{ \partial_{\varphi}^{2} - \frac{1}{\varphi^{2}} \left[\frac{q}{2} \left( \frac{q}{2} -1 \right) + \epsilon k^{2} \right] \right\} 
\label{H1 varphi}
\end{equation}
is (essentially) self-adjoint \cite{R-S} if
\begin{equation}
\frac{q}{2} \left(\frac{q}{2} -1 \right) + \epsilon k^{2} \geq \frac{3}{4} \quad \Rightarrow \quad \epsilon k^{2} \geq \frac{3}{4} - \frac{q}{2} \left(\frac{q}{2} -1\right) \quad \Rightarrow 
\begin{cases}
2 |k| \geq \sqrt{-q^{2} + 2q + 3} \quad; \, \mathrm{for} \quad \epsilon =1
\\
2 |k| \leq \sqrt{q^{2} - 2q  - 3} \quad;\,  \mathrm{for} \quad \epsilon =- 1
\end{cases}
\quad,
\end{equation}
and then we must have $-1 \leq q \leq 3$ if $\epsilon=1$ and $q \leq -1$ or $3 \geq q$ if $\epsilon=-1$ for this interval to be satisfied. To find a restriction over $p$, let us apply the same strategy as the case $\alpha = 1/3$. 

Equation (\ref{ODE X case1}) is a second-order Euler equation, and the solutions are given by (\ref{Y case 1/3}). Then, the general solutions of the eigenvalue equation (\ref{PDE}) for $\alpha = 1$ take the form
\begin{equation}
\Psi^{(i)}_{\pm} = \int_{\mathcal{K}} A(k) Y^{(i)}(\varphi,k) b^{\mu \pm \nu} dk \quad,
\label{general solution case 1}
\end{equation}
with $A(k)$ being the normalization function, and
\begin{equation}
\mu =\frac{1-p}{2} \quad ;\qquad \nu =\sqrt{\mu ^{2} + \epsilon \varpi k^{2}}\quad .
\label{mu, nu relations case 2}
\end{equation}
The solutions $Y(\varphi)$ are again combination of polynomials and modified Bessel functions, and they depend on $k$. This solution (\ref{general solution case 1}) has a similar form as (\ref{general solution case 1/3}), hence we can recall the conclusion obtained in the case $\alpha = 1/3$. The conclusion is: 
\begin{enumerate}
	\item[1.] If $p=1$ and $\varpi > 0$, then there are two possibilities:
		\begin{enumerate}
			\item[i.]  For $ q\leq -1$ and $q \geq 3$ the Hamiltonian ($\ref{Hamiltonian operator}$) is already self-adjoint.
			
			\item[ii.] It is not self-adjoint otherwise, that is $-1 <q < 3$, thought it has self-adjoint extensions.
		\end{enumerate}
	
	\item[2.] If $p=1$ and $\varpi < 0$, we have similar possibilities:
		\begin{enumerate}
			\item[i.]  For $-1 \leq q\leq 3$ the Hamiltonian ($\ref{Hamiltonian operator}$) is already self-adjoint.
			
			\item[ii.] For $q < -1$ or $q > 3$ it is not self-adjoint, but it has self-adjoint extensions.
		\end{enumerate}
\end{enumerate}

However, in principle, these are the conditions for the self-adjointness of operator $\hat{H}_{1}$ defined earlier. As a matter of fact, they are also valid for $\hat{H}$, the difference are in the boundary conditions of the extension, since we have different domains. 

\subsection{Comments}

As we pointed out earlier, there is a clear similarity between the radiative case and the stiff matter one. Perhaps the most noticible difference is that the r\^ole of the scalar field and the scalar factor exchange as we change the matter field. And, althought we did not calculate explicitly the number of possible self-adjoint extensions, for both cases we estimate that there are an infinty number of extesions, because of the superposition factor $A(k)$ as mentioned before. In fact, we can find the deficiency indices via von Neumann method (see Appendix \ref{Anexo A}), but it is a lenghty calculation compared to the method used to determine the cases which have extensions or the cases which the operator is already self-adjoint. Nevertherless, besides the number of extensions, the von Newmann method also gives the restrictions that we obtained over $p$ and $q$ for self-adjointness. 

Moreover, it is important to notice the constant of separation $k$, over which we construct the wave packets, is not completely arbitrary, it depends on the ordering factor $p$ for the case $\alpha = 1/3$ and $q$ for the case $\alpha = 1$. For example, if $\varpi <0$ in the case of the radiative matter, we must have 
\begin{equation}
\mathcal{K} = \left( - \infty, -\frac{\sqrt{-p^{2} + 2p +3}}{2} \, \right] \cup \left[ \, \frac{\sqrt{-p^{2} + 2p +3}}{2}, \infty \right) \quad.
\end{equation}
Also, remember that we can only adopt the Friedrich's extension for the case $\varpi <0$ since the Hamiltonian operator is bounded from below only in this case, both for $\alpha =1/3$ and $\alpha = 1$.

\section{Conclusion}

In this paper we have analysed the problem of self-adjointness of the Hamiltonian
operator retrieved from a Wheeler-DeWitt equation coming from the Brans-Dicke theory. We quantised a homogeneous and isotropic model with a scalar-tensor field with a non-minimal coupling, using also a generic perfect fluid matter inducing a variable for time, extending and improving the results from our previous work \cite{Ours}, where a similar analysis has been made for the minimally coupled case. As expected, we  were able to reproduce the conditions for self-adjointness obtained for a conformal matter field ($\alpha = 1/3$). Therefore, we have shown that a conformal transformation does not change the self-adjoint properties of the Hamiltonian operator. This answers one of the questions proposed in Ref. \cite{Indiano}, about the equivalence of Einstein and Jordan frames with radiative matter content. We also extended the result for the stiff matter case $\alpha=1$. 

As far as our novel contribution is concerned, we concluded that for each factor ordering considered in the quantisation, there is a particular measure in which the Hamiltonian operator becomes symmetric (Hermitian). We were able to assure the self-adjointness or the existence of self-adjoint extensions for a generic matter fluid. For the cases where there are self-adjoint extensions and the Hamiltonian operator is positive and bounded from bellow ($\varpi < 0$), we determined that the Friedrich's extension is the most natural choice, due to the nice properties coming from the fact that it is related to the associated form of the operator, such as the preservation of the ground state energy. With that in mind, we had to find out for which cases the Hamiltonian in question allows extensions or is already self-adjoint. The answer is given in terms of conditions over the ordering factor $p$ and $q$ and to obtain them we must solve the eigenvalue equation. Nevertherless, the equation is analiticaly solvable only for the cases of a radiative matter and a stiff matter. Only for those cases, we were able to find the conditions over the ordering factors and determine the self-adjointness of the operator. This also imposes a restriction on the constant of separation $k$, as obtained in \cite{Indiano} for a case with a specific ordering. However, there are cases that admits extensions, but the Hamiltonian operator is not bounded fom below, the case where $\varpi>0$. For these cases there is not a privileged extension.

We did not calculated the deficiency indices explicitly, instead, we use the similarity with the Laplacian operator and the Von Newmann method to verify the conditions for self-adjointness of the operator.
This present work is a mathematical description of the problem, in which we carefully studied the Hamiltonian operator of our theory. For future works, we intend to solve the Scr\"odinger-like equation and present some predictions, in order to produce a wider description of the quantum model of the Brans-Dicke theory, via canonical quantisation.

\appendix

\section{Calculation via deficiency indices}
\label{Anexo A}

Let us use the von Neumann method in the first case from the beginning to compare the results. With this we mean that we will calculate explicity the solutions of \eqref{PDE} for $\alpha =1/3$. In this case equation ($\ref{PDE}$) becomes,
\begin{equation}
b^{2}\left[ \frac{\ddot{X}}{X}+\frac{p}{b}\frac{\dot{X}}{X}\right]
-\varpi \varphi \left[ \varphi \frac{Y^{\prime \prime }}{Y}+q\frac{Y^{\prime }}{Y} \right] = -\eta b^{2} \quad,
\label{PDE Apendix}
\end{equation}%
from where we obtain the system of second-order ordinary differential \eqref{OED b case 1/3} and \eqref{OED varphi case 1/3}. The solutions of equation \eqref{OED b case 1/3} are,
\begin{equation}
X^{(1)} (b) = b^{\mu } H_{\nu }^{(1) }\left( \sqrt{\eta} \, b \right)  \quad; \quad X^{(2)} (b) = H_{\nu }^{(2) }\left( \sqrt{\eta } \, b\right) \quad,
\label{X solutions Apendix}
\end{equation}
where $H_{\nu }^{\left( 1,2\right) }$ are the Hankel functions and 
\begin{equation}
\mu =\frac{1-p}{2} \quad;\qquad \nu =\sqrt{\mu ^{2}+ \epsilon k^{2}} \quad.
\label{mu, nu relations}
\end{equation}%
The solutions for $\varphi$ of equation \eqref{OED varphi case 1/3} are the same as before, they are given by \eqref{Y case 1/3}. Then, the general solution of \eqref{PDE Apendix}, for each $\eta =\pm i$, is the linear combination of the particular solutions
\begin{eqnarray}
\Psi _{\pm }^{\left( 1\right) }\left( b,\varphi\right)  &=& \int_{\mathcal{K}} A^{(1)}\left( k\right) b^{\mu }H_{\nu }^{\left( 1\right) }\left( \sqrt{\eta 
} \, b\right) \varphi ^{\sigma +\rho }dk \quad ,
\label{general solution 1} \\
\Psi _{\pm }^{\left( 2\right) }\left( b,\varphi\right)  &=& \int_{\mathcal{K}} A^{(2)}\left( k\right) b^{\mu }H_{\nu }^{\left( 1\right) }\left( \sqrt{\eta 
} \, b\right) \varphi ^{\sigma -\rho }dk  \quad ,
\label{general solution 2} \\
\Psi _{\pm }^{\left( 3\right) }\left( b,\varphi\right)  &=& \int_{\mathcal{K}} A^{(3)}\left( k\right) b^{\mu }H_{\nu }^{\left( 2\right) }\left( \sqrt{\eta 
} \, b\right) \varphi ^{\sigma +\rho }dk  \quad ,
\label{general solution 3} \\
\Psi _{\pm }^{\left( 4\right) }\left( b,\varphi\right)  &=& \int_{\mathcal{K}} A^{(4)}\left( k\right) b^{\mu }H_{\nu }^{\left( 2\right) }\left( \sqrt{\eta 
} \, b\right) \varphi ^{\sigma -\rho }dk  \quad ,
\label{general solution 4}
\end{eqnarray}%
where $A^{(j) }\left( k\right) $ are normalisation functions
that depend only on $k \in \mathcal{K} \subset \mathbb{R}$. Now we have to determine which of these solutions
are square-integrable with respect to the measure ($\ref{measure}$). For
$\alpha = 1/3$, we have
\begin{equation}
\left\Vert \Psi _{\pm }^{(j) }\right\Vert =\int_{0}^{\infty
}\int_{0}^{\infty }\Psi _{\pm } ^{(j) } \, \overline{\Psi _{\pm } 
^{(j)}} \, b^{p} \, \varphi ^{q-2} \, dbd\varphi \quad .
\label{norm of the solutions}
\end{equation}%
Notice that for each case we can separate the integral over $b$ and $%
\varphi $:
\begin{equation}
\left\Vert \Psi _{\pm }^{(j)}\right\Vert =\iint_{\mathcal{K}} dk \, dk^{\prime }  A\left(
k\right) \overline{A\left( k^{\prime }\right) } \, \int_{0}^{\infty}db \, b \, H_{\nu }^{\left( 1,2\right)} \left( \sqrt{\eta } \, b\right) \overline{H_{\nu ^{\prime }}^{\left( 1,2\right)} \left( \sqrt{\eta } \, b\right) } \, \int_{0}^{\infty} d\varphi \, \varphi ^{\pm \left( \rho +\overline{\rho^{\prime }}\right) -1} \quad  .
\label{norm case 1}
\end{equation}
This way we can verify the square-integrability of the functions $X(b)$ and $Y(\varphi)$ separated.

As we argued earlier, changing the $\varphi $ coordinate as $\varphi \rightarrow e^{u}$, the integration over $\varphi$ diverges for any value of $q$ other than $1$, that is $\sigma = 0$, and we must have $\varpi<0$ for $\epsilon=1$ or $\varpi>0$ for $\epsilon=-1$. In this case
\begin{equation}
\int_{0}^{\infty }\varphi ^{\pm \left( \rho +\overline{\rho }\right)
	-1}d\varphi =\int_{-\infty }^{\infty }e^{\pm \left( \rho +\overline{\rho
		^{\prime }}\right) u} \, du = 2\pi \sqrt{\vert \varpi \vert}\delta \left( \pm \left\vert k\right\vert \mp \left\vert k^{\prime }\right\vert \right)\quad.
\label{int varphi Appendix}
\end{equation}
We integrate over $k$ for the solution with a positive exponent and over $%
k^{\prime }$ for the solution with the negative exponent. 

After integrating in $\varphi$ and $k$, the resulting norm is
\begin{equation}
\left\Vert \Psi _{\pm }^{(j) }\right\Vert = 2 \pi \sqrt{|\varpi|}%
\int_{0}^{\infty }\int_{\mathcal{K}} \, \left\vert A\left( k\right)
\right\vert ^{2} b \, H_{\nu }^{\left( 1,2\right) }\left( \sqrt{\eta } \, b\right) H_{\overline{\nu }}^{\left( 1,2\right) }\left( \sqrt{\eta } \, b\right) dkdb\quad .
\label{norm case 1.a}
\end{equation}
Now we have to inspect how the integral over $b$ behaves. Let us analyse the Hankel functions assymptotically. For the limit $b\rightarrow 0$ we have
\begin{equation}
H_{\nu }^{\left( 1,2\right) }\left( \sqrt{\eta } \, b\right) \sim \left( \sqrt{%
	\eta } \, b\right) ^{-\nu }  \quad.
\label{Hankel func. b<<}
\end{equation}%
That is,
\begin{equation}
b \, H_{\nu }^{\left( 1,2\right) }\left( \sqrt{\eta} \,b \right) H_{\overline{\nu }}^{\left( 1,2\right) }\left( \sqrt{\eta } \, b\right) \sim \left\vert \sqrt{\eta }\right\vert ^{2}b^{1-\nu -\bar{\nu}}=b^{1-2Re\left( \nu \right) } \quad.
\label{norm argument b<< }
\end{equation}%
Therefore, for $b<<$, the norm behaves as
\begin{eqnarray}
\left\Vert \Psi _{\pm }^{(j)}\right\Vert^{2} = \lim_{b \rightarrow 0} \left\{ 2 \pi \sqrt{|\varpi|} \int_{\mathcal{K}} \frac{\left\vert A\left( k\right) \right\vert ^{2}}
{\left( 1-Re\left( \nu \right) \right)}b^{2-\nu -\bar{\nu}}dk \right\}; .
\label{norm b<<}
\end{eqnarray}%
Considering that the superposition factor $A\left( k\right) $ is well
defined in the interval $\mathcal{K}$, the norm goes to zero only if we have $\left[ 2-2Re\left( \nu \right) \right] >0$, that is,
\begin{equation}
2-2\sqrt{\mu ^{2} + \epsilon k^{2}}> 0 \quad \Longrightarrow \quad \epsilon k^{2} < 1- \mu ^{2} \quad \Longrightarrow \quad
\begin{cases}
\left\vert k \right\vert < \sqrt{1- \mu ^{2}} \quad, \quad \epsilon = 1 \\
\left\vert k \right\vert > \sqrt{\mu ^{2}-1} \quad, \quad \epsilon = -1
\end{cases} 
\quad.
\end{equation}%
This interval only exists if we have $\mu ^{2}\geq 1$ for the case $\epsilon=1$ or $\mu ^{2}\leq 1$ for the case $\epsilon=-1$. Hence, from \eqref{mu, nu relations}, we must have $-1 \leq p \leq 3$ for $\epsilon=1$ and $p\leq-1$ and $p\geq3$ for $\epsilon=-1$ to obtain convergence. Outside these intervals, there is no convergence and the operators are self-adjoint, in agreement with whathas been found in Subsection \ref{Subsec 1/3}.

Let us verify the limit $b\rightarrow \infty $. In this case we have
\begin{equation}
H_{\nu }^{\left( 1,2\right) }\left( \sqrt{\eta } \, b\right) \sim \sqrt{\frac{2}{\pi \sqrt{\eta } \, b}}e^{\pm i\left( \sqrt{\eta } \, b-\frac{\pi \nu }{2}-\frac{\pi }{2}\right) } \quad,
\label{Hankel func. b>>}
\end{equation}%
where the $\left( \pm \right) $ on the exponential corresponds to the
Hankel functions of the first and second kind. Thus, recovering $\eta
=\left( \pm i\right) $, that is, $\sqrt{\eta}=\frac{1}{\sqrt{2}}\left(i\pm1\right)$, the norm becomes 
\begin{equation}
\left\Vert \Psi _{\pm }^{\left( i\right) }\right\Vert^{2} = \lim_{b \rightarrow \infty} 
\left\{ \left\vert \kappa \right\vert ^{2}\int_{-\infty }^{\infty }\left\vert A\left( k\right)
\right\vert ^{2}P_{k}(b) \, e^{\pm i\left[ \frac{1}{\sqrt{2}}\left( i\pm 1\right) b-%
	\frac{\pi }{2}\left( \nu -\bar{\nu}\right) \right] }dk \right\} ,
\label{norm b>>}
\end{equation}
where $P_{k}(b) $ is a polinomial of $b$ that depends on $k$ and $\kappa$ is a complex number. Therefore, we conclude that for $\eta =i$ the Hankel function of the first kind converges, as the Hankel function of the second kind diverges and for $\eta =-i$ is the opposite. Therefore, we recovered the conclusion obtained in Subsection \ref{Subsec 1/3}. A similar calculation can be done to recover the results of the case $\alpha = 1$. 

\vspace{0.7cm}

\noindent
{\bf Acknowledgements:} C.R.A., A.B.B and J.C.F. thanks CNPq (Brazil) and FAPES (Brazil) for partial financial support. PVM is supported by the grant PEst-OE/MAT/UI0212/2014. We all thank Giuseppe Dito for many hours of enlightening discussions and for many valuable suggestions.

\end{document}